\documentclass[aps,prb,twocolumn]{revtex4-2}
\usepackage{graphicx}
\usepackage{graphics}
\usepackage{amsfonts}
\usepackage{amsmath}
\usepackage{amssymb}
\usepackage{xcolor}

\begin{document}

\title{Spin transfer exchange torque in ferromagnet/ferromagnet structures
\\
made of half metals with large exchange gaps}

\author{V. Shablenko}
  \email{shablenv@email.sc.edu}

\author{Ya.\ B. Bazaliy}
  \email{bazaliy@mailbox.sc.edu}
  \thanks{corresponding author}

\affiliation{University of South Carolina, Columbia SC 29208, USA}

\date{\today}

\begin{abstract}
Spin torques in magnetic multilayers are produced by spin polarization $P$ of ferromagnetic (F) layers, and increase with growing $P$. The latter, however, cannot exceed the $P=1$ value found in half metals. We study the $P=1$ case to find what other parameters still influence spin torques in this extreme limit. It is found that the ratio of exchange gap to Fermi energy strongly affects the properties of the torque. For large values of the gap the magnitude of  exchange spin torque  exhibits a sharp peak at very small misalignment angles between magnetizations. This behavior is found to be linked to a transition between Ohmic and tunneling transport regimes through the F/F boundary.
\end{abstract}

\maketitle

\section{Introduction}
In metallic F/N/F spin valves with F denoting ferromagnetic and N normal layers the motion of itinerant electrons between the layers produces torques acting on the magnetizations of ferromagnets. In equilibrium the interlayer exchange torque (RKKY-type exchange) is produced \cite{slonczewski-tunnel:1989, erickson-hathaway_cullen:1993, slonczewski_interlayer:1993, stiles:1993}, and when electric current is passed through the valve, non-equilibrium corrections to the main torque appear \cite{slonczewski_ST:1996, berger_ST:1978, berger_ST:1986, berger_ST:1996}. If magnetizations of left and right F-layers are ${\bf M}_L$ and ${\bf M}_R$, the equilibrium torque ${\bf T}_{eq}$ is pointed perpendicular to the $({\bf M}_L$, ${\bf M}_R)$ plane. Electric current $I_e$ directed perpendicular to F/N boundaries produces two types of corrections. First, there is a change in magnitude of ${\bf T}_{eq}$ \cite{zhang_non-adiabatic:2004, chshiev:2008, tang:2009}. This correction $\Delta {\bf T}_{\perp}(I_e)$ is perpendicular to the plane formed by ${\bf M}_L$ and ${\bf M}_R$, and is alternatively called an ``out-of-plane'' or a ``field-like'' torque. Second correction $\Delta {\bf T}_{||}(I_e)$ is the Slonczewski-Berger spin-transfer torque, directed in the $({\bf M}_L$, ${\bf M}_R)$ plane \cite{slonczewski_ST:1996}. It is alternatively called an ``in-plane'' or a ``dissipation-like'' torque.

In this paper we study a system that differs from a conventional F/N/F valve in several ways.

First, we consider ferromagnets with full spin polarization $P$ of itinerant electrons,
i.e., half metals with $P = 1$. Full spin polarization can be realized in Heusler alloys that enjoy recent popularity in spintronic literature \cite{elphick:2021, hirohata:2022, parkin:2022, tavares:2023, ren:2023}. Many novel two-dimensional ferromagnets are also theoretically predicted to be half-metals \cite{ashton:2017, li:2020, hao:2021}. Since one cannot increase $P$ beyond unity, a question may be posed of whether all fully spin-polarized ferromagnets produce the same, largest possible spin torque. If not, what other material parameters influence spin torques in this regime?

\begin{figure}[b]
\center
\includegraphics[width = 0.45\textwidth]{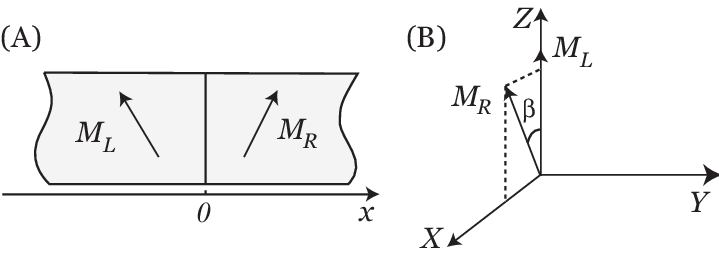}
\caption{(A) F/F device; (B) coordinates in spin space}
 \label{fig:FF_device}
\end{figure}

Second, we consider F/F structures without a normal spacer (Fig~\ref{fig:FF_device}A). It is still assumed that magnetizations of ferromagnets can be rotated independently and that the exchange interaction between the layers is produced by itinerant electrons, as opposed to the overlap of localized orbitals of atoms near the boundary. To justify such assumption one can imagine that F-layers are in fact separated by an ultra-thin normal layer. If the thickness of such layer is smaller than the itinerant electron wavelength but larger than the extent of localized orbitals, the F/F idealization is possible.

In a naive picture, at $P=1$ an electron passing through an F/F structure has its spin parallel to ${\bf M}_L$ everywhere on the left and parallel to ${\bf M}_R$ everywhere on the right of the boundary, so an instant spin rotation should take place at the boundary. Explaining how it happens is our first task.

It is also clear that as angle $\beta$ between ${\bf M}_L$ and ${\bf M}_R$ (defined in Fig.~\ref{fig:FF_device}B) increases, it becomes progressively harder for an electron to cross the F/F boundary. For $P =1$ all electrons are fully reflected in the anti-parallel configuration, $\beta = \pi$. Conversely, in a parallel configuration, $\beta =0 $, the boundary is transparent. As a result, a fully spin-polarized F/F contact will operate in an Ohmic regime for $\beta \approx 0$, and in a tunneling regime for $\beta \approx \pi$, even in the absence of any real tunnel barrier at the boundary. The presence of a crossover between two regimes differentiates the F/F case from that of F/N/F structures, where both F/N boundaries may remain in the Ohmic regime at any $\beta$, even in the case of full spin polarization.

In this work we treat the case of identical F-layers. This situation already demonstrates the features brought about by our third assumption: ferromagnetic exchange gap $\Delta$ greatly exceeds the Fermi energy $\epsilon_F$ of itinerant electrons. We expect the limit of $\Delta \gg \epsilon_F$ to exhibit the most pronounced features of the fully spin-polarized regime. The presence of small parameter $\epsilon_F/\Delta \ll 1$ allows us to derive analytic expressions, which is the technical achievement of this work. Using them, we find that many characteristics of F/F valves exhibit strong variations at very small angles $\beta$. In particular, our calculations show that the crossover from Ohmic to tunneling behaviour happens at an angle $\beta_c \sim (\epsilon_F/\Delta)^{1/4} \ll 1$. Furthermore, the same characteristic angle shows up in the equilibrium exchange torque angular dependence $T_{eq}(\beta)$ that exhibits a sharp peak at $\beta_c$. We explain the physics behind this high sensitivity to small angular deviations and discuss how rapid non-linear changes of conductance and torques at small angles may open new possibilities for building efficient devices based of fully spin-polarized ferromagnets.

\section{Model of ferromagnets}
In line with the majority of spin-transport studies, we employ a single-electron approximation to describe the F/F valve. Namely, Stoner model \cite{coey_textbook_stoner_model} with exchange-split bands is used. With spin quantization axis $Z$ chosen along the magnetization $\bf M$,  one-electron Hamiltonian reads
\begin{equation} \label{eq:H}
\hat H = \left| \begin{array}{cc}
- \frac{\hbar^2 \nabla^2}{2m} + eV(r) & 0 \\
0 & - \frac{\hbar^2 \nabla^2}{2m} + \Delta + eV(r)
\end{array} \right|
\end{equation}
where $V(r)$ is the electric potential and $e < 0$ is the electron charge. In a bulk ferromagnet we can set $V = 0$ and get the band energies
\begin{equation}
\epsilon_{\uparrow} = \frac{\hbar^2 k^2}{2 m}, \qquad
\epsilon_{\downarrow} = \frac{\hbar^2 k^2}{2 m} + \Delta.
\end{equation}
A ferromagnet is fully spin-polarized when
\begin{equation}
\Delta > \epsilon_F = \frac{\hbar^2 k_F^2}{2 m}
\end{equation}
and all electrons reside in the lower band. Electron plane waves in the bulk are then described by spinors
\begin{equation}
\left( \begin{array}{c}
\psi_{\uparrow} \\ \psi_{\downarrow}
\end{array} \right) =
\left( \begin{array}{c}
1 \\ 0
\end{array} \right) e^{i {\bf k} \cdot {\bf r}}.
\end{equation}

Note that Stoner model does not include spin-orbit interactions, thus real and spin spaces are completely decoupled and can be arbitrarily rotated with respect to each other. We reserve $(x,y,z)$ coordinates for real space, with $x$ being perpendicular to the boundary. Spin space coordinates are denoted by capital letters $(X,Y,Z)$.

\begin{figure}[t]
\center
\includegraphics[width = 0.4\textwidth]{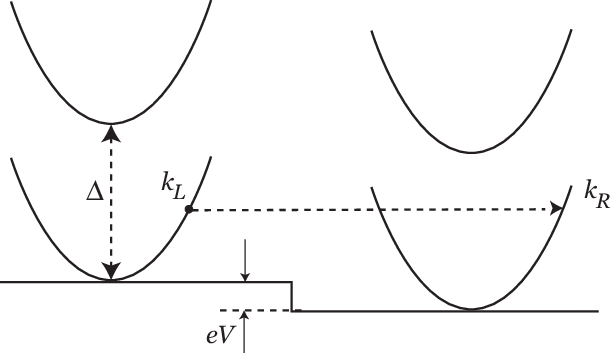}
\caption{Energy diagram for electrons crossing the F/F boundary. Ferromagnets on both sides of the structure are assumed to be identical.}
 \label{fig:energy_diagram}
\end{figure}

\section{Single-electron contributions to currents through the boundary}
\subsection{Transmission and reflection amplitudes}

In an F/F structure electron coming from the left is partially reflected from the boundary when $\beta \neq 0$. The reflection process is described by the standard matching of wave functions and their derivatives at the interface. Electron transmission is illustrated by the energy diagram in Fig.~\ref{fig:energy_diagram}. Wave function matching can be achieved only if nonzero spin-down components appear in the form of evanescent waves, decaying far away from the boundary. For example, for an electron incoming from the left ($L \to R$ process) wave function is given by spinors
\begin{equation}\label{eq:incoming_spinor_L-R}
\left( \begin{array}{c}
\psi_{\uparrow} \\ \psi_{\downarrow}
\end{array} \right)_L =
\left( \begin{array}{c}
e^{i k_L x} + r e^{-i k_L x} \\ d_L e^{\kappa_L x}
\end{array} \right)
, \quad (x < 0),
\end{equation}
and
\begin{equation}\label{eq:transmitted_spinor_L-R}
\left( \begin{array}{c}
\psi_{\uparrow} \\ \psi_{\downarrow}
\end{array} \right)_R =
\left( \begin{array}{c}
t e^{i k_R x}  \\ d_R e^{-\kappa_R x}
\end{array} \right)
,  \quad (x > 0).
\end{equation}
The presence of spin-down components describes the ``instantaneous rotation'' of spin upon boundary crossing. This rotation is, in fact, happening on a lengthscale set by the evanescent waves.

Appendix~\ref{appsec:single-electron_currents} gives the details of derivation for reflection and transmission amplitudes. They are expressed in terms of perpendicular to the boundary components $k_{L}$ and $k_{R}$ of wave vectors. In a general case with possible voltage drop on the boundary, energy conservation in transmission process dictates relations
\begin{eqnarray}
\label{eq:kappa-k-delta}
  k_{L}^2 + \kappa_L^2 &=& k_{R}^2 + \kappa_R^2 = k_{\Delta}^2 \ ,
\\
\label{eq:kL-kR}
  k_{L}^2 + k_V^2 &=& k_{R}^2.
\end{eqnarray}
where $k_{\Delta}^2 = 2 m \Delta/\hbar^2$, $k_V^2 = 2 m e V/\hbar^2$. For a $L \to R$ process we find
\begin{equation}\label{eq:reflection_through_k's}
  \begin{gathered}
r = - \frac{(\kappa_R + i k_L)(\kappa_L - i k_R) s^2
- i (\kappa_L + \kappa_R)(k_R -k_L)c^2}{D} ,
\\
d_L = - \frac{2 i k_L (\kappa_R + i k_R) \ s \ c}{D} \ ,
  \end{gathered}
\end{equation}
where
\begin{equation}\label{eq:D-s-c}
  \begin{gathered}
s = \sin(\beta/2), \quad c = \cos(\beta/2) \ ,
\\
D = (\kappa_R - i k_L)(\kappa_L - i k_R) s^2 - i (\kappa_L + \kappa_R)(k_R + k_L)c^2 \ .
  \end{gathered}
\end{equation}
Amplitudes $r$ and $d_L$ completely determine the wave function on the left of the boundary.

\subsection{Single-electron currents}
We assume all currents to flow perpendicular to the boundary, i.e., in the $x$ direction in real space. In terms of spinors $\psi_{\alpha}(x)$, ($\alpha = \uparrow, \downarrow$) single-electron electric current is given by the standard expression \cite{schiff_QMtextbook}
$$
j^{(1)}_e(x) = e \frac{\hbar}{2 m} \big[
\psi_{\alpha}^* (-i\partial_x \psi_{\alpha}) +  c.c.
\big],
$$
where $e <0$ is electron charge, ``$c.c.$'' is the complex conjugate, and summation over repeated indices $\alpha$ is implied. It is sometimes beneficial to discuss particle current $j_n = j_e/e$ instead of electric current.

Single-electron spin current is given by \cite{erickson-hathaway_cullen:1993}
\begin{equation}\label{eq:spin-current_definition}
j^{(1)}_{sa} = \left(\frac{\hbar}{2}\right) \frac{\hbar}{2 m} \big[
\psi_{\alpha}^* \sigma^a_{\alpha\beta} (-i\partial_x \psi_{\beta}) +  c.c.
\big].
\end{equation}
where $a = X,Y,Z$, $\sigma^a$ are the Pauli matrices, and summation over repeated Greek indices is again implied.

Single-electron currents can be re-written in terms of reflection amplitudes. For the $L \to R$ process particle current is given by
\begin{equation}\label{eq:je_through r}
j^{(1)}_n = \frac{\hbar k_L}{m} \big( 1 - |r|^2) \ .
\end{equation}
Spin current for the $L \to R$  process is obtained in the region $x < 0$ by substituting spinor (\ref{eq:incoming_spinor_L-R}) into definition (\ref{eq:spin-current_definition}). For our purposes we will only need the $y$-component of spin current, for which we obtain an expression
\begin{equation}\label{jY_through_r_and_dL}
j^{(1)}_{sY}(x) = -\frac{\hbar^2}{4 m} \big[
(\kappa_L + i k_L)(d_L + r d_L^{*}) e^{i k_L x} +  c. c.
\big].
\end{equation}

\section{Exchange interaction in equilibrium}

Spin torques acting on the magnetizations of F-layers are related to total spin currents $j_{sa}$ ($a = X,Y,Z$) entering and leaving each layer \cite{slonczewski_ST:1996}. Torque acting on the left layer has components (see Appendix \ref{appsec:jsY-TLY})
\begin{equation}\label{eq:T_through_Is}
T_{La} = -j_{sa}(0) + j_{sa}(-\infty) \ .
\end{equation}
Total spin currents are the sums of contributions from electrons moving in $L \to R$ and $R \to L$ directions
$$
j_{sa} = \sum_{L \to R} j^{(1)}_{sa}({\bf k}) + \sum_{R \to L} j^{(1)}_{sa}({\bf k}) \ .
$$
Deep in the ferromagnet spin current is directed along the magnetization and completely determined by the total particle current
\begin{equation}\label{eq:jsa(infty)}
j_{sa}(-\infty) = (\hbar/2) P j_n \delta_{Za} \ .
\end{equation}
This relation holds in any ferromagnetic material. However, in partially spin-polarized materials it requires spin-relaxation processes to be included in the model \cite{rashba_diffusive:2000, rashba_diffusive:2002, spin-current_book}. In the present case of full spin polarization we have a much simpler picture with relation (\ref{eq:jsa(infty)}) being already enforced by the single-electron Hamiltonian (\ref{eq:H}), as one can see from spinor expression (\ref{eq:incoming_spinor_L-R}).

In this section we wish to find spin torque in equilibrium, where $j_{sa}(-\infty) = 0$ and we only need to evaluate the sum for $j_{sa}(0)$. In the state of equilibrium equal numbers of electrons cross the boundary in two directions. There is no potential drop on the boundary, and any $L \to R$ moving electron with perpendicular momentum $k_L$ has a partner moving in the $R \to L$ direction with $k_R = - k_L$ and same values of $\epsilon$ and ${\bf k}_{||}$. Similar to the case of spin torque in tunneling contact \cite{slonczewski-tunnel:1989, chshiev:2008}, a direct computation shows that contributions of partners to $j_{sX}(0)$ and $j_{sZ}(0)$ are opposite and cancel each other, leading to the expected absence of Slonczewski-Berger torque $\Delta{\bf T}_{||}$. In contrast, partner's contributions to $j_{sY}(0)$ are equal and add up, doubling the $L \to R$ sum
\begin{equation}\label{eq:TLY}
j_{sY}(0) = 2 \sum_{L \to R} j^{(1)}_{sY}({\bf k},0) \ ,
\end{equation}
where we use notation
$$
j^{(1)}_{sY}({\bf k},0) = \left. j^{(1)}_{sY}({\bf k}) \right|_{x = 0} \ .
$$
$Y$-component of spin current produces the only non-zero component ${T}_{LY}$ of the equilibrium exchange torque.

The sum (\ref{eq:TLY}) is performed over all right-moving electrons inside the Fermi sphere, i.e., over the region $\Omega$ defined by $|{\bf k}|\leq k_F$ and $k_L \geq 0$
$$
j_{sY}(0) = 2 \sum_{{\bf k} \in \Omega} j^{(1)}_{sY}(k_L,0)
= 2 \int_{\Omega} j^{(1)}_{sY}(k_L,0) \frac{d^3k}{(2\pi)^3} \ .
$$
In spherical coordinates with $k = |{\bf k}|$ and polar angle $\theta$ measured from the $x$-axis one has $k_L = k \cos\theta$ and the integral acquires a form
\begin{eqnarray}
\nonumber
j_{sY}(0) &=& 2 \int_0^{k_F} \int_0^{\pi/2}\int_0^{2\pi}
j_{sY}^{(1)}(k \cos\theta,0)
\frac{k^2 dk \sin\theta d\theta d\phi}{(2\pi)^3}
\\
\label{eq:Iy_total_exact}
&=& \frac{4\pi}{(2\pi)^3}  \int_0^{k_F} k^2 \int_0^1
j_{sY}^{(1)}(\xi k,0) \, d\xi \ .
\end{eqnarray}
We were able to compute this integral analytically in the limit of large exchange gap, $\Delta \gg \epsilon_F$ or eqivalently $k_{\Delta} \gg k_F \geq k_L$. This limit is discussed in Appendix~\ref{appsec:single-electron_currents}, where amplitudes (\ref{eq:reflection_through_k's}) are approximated, and formula (\ref{jY_through_r_and_dL}) acquires a form
\begin{equation}\label{eq:jsy_at_x=0}
j^{(1)}_{sY}(k_L,0) = - \left( \frac{\hbar^2}{4 m} \right) \frac{16 k_L^2 k_{\Delta} s \, c}{k_{\Delta}^2 s^4 + 16 k_L^2 c^2}
\end{equation}
Using it, we first compute the angle integral
\begin{eqnarray*}
&& \int_0^1  j_{sY}^{(1)}(\xi k,0) \, d\xi =
- \left( \frac{\hbar^2}{4 m} \right) k_{\Delta} \frac{s}{c}
\int_0^1 \frac{\xi^2 d\xi}{\nu^2 + \xi^2}
\\
&& = - \left( \frac{\hbar^2}{4 m} \right) k_{\Delta} \frac{s}{c}
\left( 1 - \nu \arctan\frac{1}{\nu} \right) \ ,
\end{eqnarray*}
where $\nu = \nu(k,\beta) = k_c(\beta)/k$ with $k_c(\beta) = k_{\Delta} s^2/4 c$. Substituting this result into formula (\ref{eq:Iy_total_exact}) we get
\begin{equation}\label{eq:jsY_with_coefficient}
j_{sY}(0) = - \left(
\frac{4\pi}{(2\pi)^3} \frac{\hbar^2}{4 m} k_{\Delta} s c
\right) I(\beta) \ ,
\end{equation}
where integral $I(\beta)$ can be taken analytically as well
\begin{eqnarray}
\nonumber
I(\beta) &=&  \int_0^{k_F} k^2 \left( 1 - \frac{k_c(\beta)}{k} \arctan\frac{k}{k_c(\beta)} \right) dk
\\
\nonumber
& = & \frac{k_F^3}{3} - k_c^3 \int_0^{k_F/k_c} \xi \, \arctan\xi \ d\xi
\\
\nonumber
&=& \frac{k_F^3}{3} - \frac{k_c^3}{2} \left\{
\left[\left(\frac{k_F}{k_c}\right)^2 + 1 \right] \arctan \frac{k_F}{k_c}
- \frac{k_F}{k_c}
\right\} \ .
\end{eqnarray}
The result can be expressed through a dimensionless parameter
\begin{equation}\label{eq:nu_definition}
\nu_F(\beta) = \nu(k_F,\beta) = \frac{k_{\Delta} s^2}{4 k_F c}
\end{equation}
as
\begin{eqnarray}
\nonumber
I(\beta) &=& k_F^3 \phi_s\big( \nu_F(\beta) \big) \ ,
\\
\label{eq:definition_phi_s}
\phi_s(\nu_F) &=& \frac{1}{3} - \frac{1}{2}
\left(
\nu_F (1 + \nu_F^2) \arctan\frac{1}{\nu_F} - \nu_F^2
\right).
\end{eqnarray}
Substituting expression (\ref{eq:definition_phi_s}) into formula (\ref{eq:jsY_with_coefficient}), and then into (\ref{eq:T_through_Is}) we find the exchange torque
\begin{equation}\label{eq:TLY_final_approximation}
T_{LY} =  - j_{sY}(0)
= \frac{\pi \hbar^2 k_{\Delta}k_F^3}{(2\pi)^3 m}  \frac{s}{c} \, \phi_s(\nu_F) \ .
\end{equation}
As explained in Appendix~\ref{appsec:jsY-TLY}, positivity of $T_{LY}$ corresponds to ferromagnetic exchange between the F-layers. That appendix also explains that the assumption made in our model about majority electron's spins being parallel to $\bf M$ is not crucial: had we assumed majority spins being antiparallel to $\bf M$, the sign of exchange interaction would remain ferromagnetic.

Factor $k_{\Delta} k_F^3$ in expression (\ref{eq:TLY_final_approximation}) can be understood as the number of electrons participating in the process $n_e \sim k_F^3$ multiplied by $k_{\Delta}$ that quantifies the strength of exchange interaction between itinerant electrons and local magnetization.

\begin{figure}[t]
\center
\includegraphics[width = 0.48\textwidth]{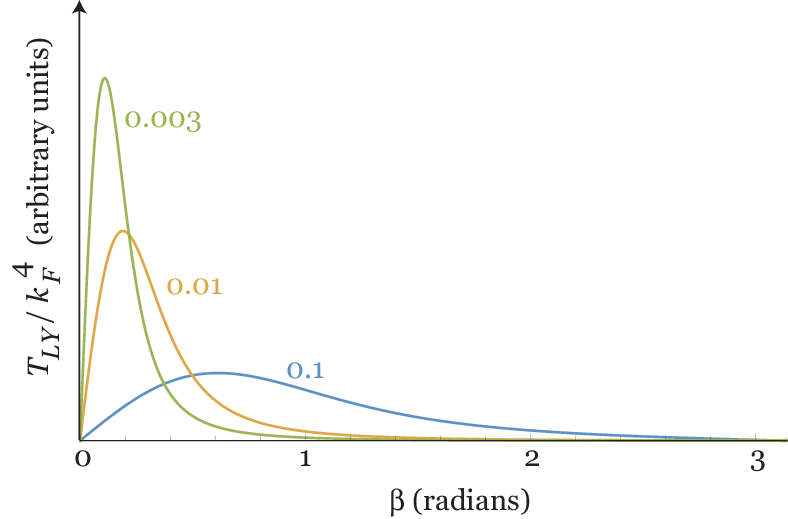}
\caption{Angular dependence of normalized exchange torque $T_{LY}(\beta)/k_F^4$. Numbers near curves give the values of $k_F/k_{\Delta}$.}
 \label{fig:I_vs_angle}
\end{figure}

Typical angular dependencies of exchange torque $T_{LY}(\beta)$ are shown in Fig.~\ref{fig:I_vs_angle} (for reasons that will become clear below, this figure shows torques normalized by $k_F^4$). Torque graphs exhibit tall narrow peaks---in contrast with broad angular dependencies of torques found in conventional spin valves, where only quadratic and biquadratic exchange components are important \cite{slonczewski_interlayer:1993, tang:2009}. Sharp variations of torque at small angles mathematically follow from the form of function $\nu_F(\beta)$. As angle $\beta$ is swept from zero to $\pi$, the value of $\nu_F$ increases from zero to infinity. In the limit $k_F/k_{\Delta} \ll 1$ the crossover between small and large values of $\nu_F$ happens in a narrow range of angles. Namely, $\nu_F$ changes from 0.1 to 10 in the interval $0.3 \beta_c < \beta < 3 \beta_c$, where the characteristic angle
\begin{equation}\label{beta_c_definition}
\beta_c \approx 4 \sqrt{k_F/k_{\Delta}} \ll 1
\end{equation}
is an approximate solution of equation $\nu_F(\beta) = 1$ in that limit. Locations of peaks in Fig.~\ref{fig:I_vs_angle} are closely related to $\beta_c$. Equation $d T_{LY}/d\beta = 0$ is solved in Appendix~\ref{appsec:approximations_equilibrium_exchange} and gives the peak position
\begin{equation}\label{eq:beta_max}
\beta_{max} = 2 \sqrt{\frac{8 k_F}{3 \pi k_{\Delta}}} \approx 0.46 \beta_c \ .
\end{equation}
The width of the peak is of order $\beta_c$ as well.

The height of the peak, calculated in Appendix~\ref{appsec:approximations_equilibrium_exchange}, is
\begin{eqnarray}
\nonumber
T_{LY}(\beta_{max}) &=& \left( \frac{\pi \hbar^2 k_F^4}{(2\pi)^3 m} \right) \sqrt{\frac{8 k_{\Delta}}{3 \pi k_{F}}} \ \phi_s(2/3\pi)
\\
\label{eq:T_max}
& \approx & \left( \frac{\pi \hbar^2 k_F^4}{(2\pi)^3 m} \right)  0.19 \sqrt{\frac{k_{\Delta}}{k_{F}}}  \ .
\end{eqnarray}

If $k_F$ is kept constant, the peaks grow higher and more narrow as the ratio $k_F/k_{\Delta}$ is decreased (Fig~\ref{fig:I_vs_angle}). However, the product of peak height by its width remains constant. Physical meaning of this fact is revealed by calculating of exchange energy $E_{ex}$. When magnetizations make angle $\beta$ with each other, the energy of the system is higher than at $\beta = 0$ by
$$
E_{ex}(\beta) = \int_0^{\beta} T_{LY}(\beta') d\beta' \ .
$$
Total exchange energy is given by an integral
\begin{equation}\label{eq:Eex_integral}
E_{ex}^{tot} = \int_0^{\pi} T_{LY}(\beta) d\beta \ .
\end{equation}
Calculations detailed in Appendix~\ref{appsec:approximations_equilibrium_exchange} result in
\begin{equation}\label{eq:Eex_approx}
E_{ex}^{tot} \approx \frac{\pi^2}{4(2\pi)^3}  \frac{\hbar^2 k_F^4}{m} \ .
\end{equation}
This shows that exchange energy becomes independent of the ferromagnetic gap value when the latter is infinitely increased---in that respect all ``infinitely-polarized'' ferromagnets are equal. Instead of changing the magnitude of $E_{ex}^{tot}$, large gap leads to the evolution of the shape of function $E_{ex}(\beta)$: the latter rapidly increases from zero to almost full $E_{ex}^{tot}$ in the vicinity of $\beta_c$, and later remains nearly constant all the way to $\beta = \pi$.

\section{Boundary conductance}

As discussed in the introduction, an important feature of the F/F system is the transition from Ohmic to tunneling conductance regime upon the increase of angle $\beta$. A model describing both regimes on equal footing necessarily needs to take into account non-isotropic deviations of electron distribution function $n({\bf k},x)$ from equilibrium. Indeed, for $\beta = 0$ physical boundary between identical ferromagnets  disappears,  and particle current at $x = 0$ assumes its bulk form
$$
j_n(0) = \int \frac{\hbar k}{m} n({\bf k},0) \frac{d^3k}{(2\pi)^3} \ .
$$
Clearly, only non-isotropic distribution function $n({\bf k},0)$ can ensure non-zero value of the integral on the right.

At non-zero $\beta$ both the non-isotropic nature of $n$, and the potential jump $eV$ contribute to the electric current through the boundary. In a fully self-consistent approach, electron distribution function $n({\bf k},x)$ is allowed to have a jump at $x = 0$ with $n({\bf k},0_-) = n_L({\bf k})$ and $n({\bf k},0_+) = n_R({\bf k})$. Away from the boundary, the distribution gradually changes, until acquiring its bulk, current-carrying form at $x \to \pm\infty$. This gradual change is governed by the Boltzmann equation. Self consistency is achieved when the potential jump $eV$ and functions $n_{L,R}$ are such that electric current through the boundary equals the bulk current.

\begin{figure}[t]
\center
\includegraphics[width = 0.45\textwidth]{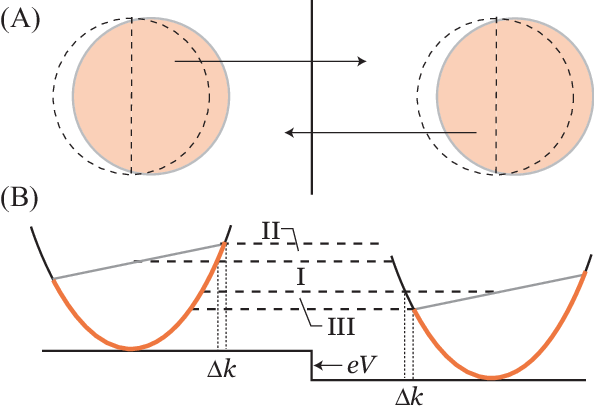}
\caption{Model for electric current calculation. (A) Electron distribution in $k$-space. Distributions on both sides of the boundary are shown. Filled electron states of the ``shifted Fermi spheres'' are shown in red color; equilibrium Fermi spheres are shown by dashed lines. Vertical dashed lines separate the $k$-states of right-moving and left-moving electrons. Current produced by $L \to R$ moving electrons is partially compensated by  $R \to L$ moving electrons. (B) Energy diagram. Electrons contained in energy intervals (I), (II), and (III) are uncompensated and contribute to the total current.}
 \label{fig:electric_current_diagram_maintext}
\end{figure}

We will use an approximate description that restricts $n({\bf k},x)$ to a pre-determined form specified by just a few parameters, and imposes self-consistency condition on those parameters. Namely, we will assume that electron distribution functions in F-layers are space-independent and given by the simplest current-carrying solution of the bulk Boltzmann equation with electrons occupying states inside a ``shifted Fermi surface'' (Fig.~\ref{fig:electric_current_diagram_maintext}A) \cite{kittel_intro_solid_state, ashcroft-mermin_solid_state}.  Two parameters, the shift $\Delta k$ and the boundary potential jump $eV$, determine both bulk and boundary currents $j_e$. Condition for those currents to be equal establishes a relation between $\Delta k$ and $eV$, and ultimately determines the boundary conductance $G = j_e/V$.

Calculation is performed in two steps. First, electric current through the boundary is calculated for arbitrary $\Delta k$ and $eV$, without imposing self-consistency condition. The result has ``Ohmic'' and ``tunneling'' contributions, as illustrated by the energy diagram in Fig.~\ref{fig:electric_current_diagram_maintext}B. Ohmic contributions are produced by the non-isotropic parts of $n({\bf k})$, represented by zones (II) and (III) in the figure. Tunneling contribution is produced by zone (I). The whole approach is conceptually simple but involves many technical steps detailed in Appendix~\ref{appsec:electric_current_non_self-consistent}, where approximate single-electron contributions found in Appendix~\ref{appsec:single-electron_currents} are summed over zones (I), (II), and (III). The result, expressed through dimensionless $\nu_F$, reads
\begin{eqnarray}
\label{eq:jn_total}
j_n &=&   \frac{1}{(2\pi)^3} \frac{\hbar k_F^4}{m}
\left[
\frac{\pi\phi_1}{2} \frac{eV}{\epsilon_F} + 4\pi \phi_2 \frac{\Delta k}{k_F}
\right]
\end{eqnarray}
with
\begin{eqnarray}
\label{eq:phi1}
\phi_1 &=& 1 - \nu_F^2 \ln\left(1 + \frac{1}{\nu_F^2}\right) \ ,
\\
\label{eq:phi2}
\phi_2 &=& \frac{1}{3} - \nu_F^2 \left( 1 - \nu_F \, \arctan \frac{1}{\nu_F}  \right).
\end{eqnarray}

Appendix~\ref{appsec:self-consistent_eV} derives the self-consistency requirement on $\Delta k$ and $eV$, imposed by the equality of currents through the boundary and in the bulk. It reads
\begin{equation}\label{eq:self-consistency}
\frac{\Delta k}{k_F} = \frac{\phi_1}{8(1/3 - \phi_2)} \frac{eV}{\epsilon_F} \ .
\end{equation}
Using this relation in formula (\ref{eq:jn_total}) one can express current in terms of $eV$ alone
$$
j_e = \frac{\pi}{(2 \pi)^3} \frac{e^2 k_F^2}{\hbar} \frac{\phi_1}{1 - 3\phi_2} V
= G(\beta) V \ ,
$$
and find the conductance of the boundary
\begin{eqnarray}
\label{eq:boundary conductance}
G(\beta) &=& G_0 \frac{\phi_1}{1 - 3\phi_2} \ ,
\\
\label{eq:G0}
G_0 &=& \frac{\pi}{(2 \pi)^3} \frac{e^2 k_F^2}{\hbar} \ .
\end{eqnarray}
As expected, conductance diverges when $\beta \to 0$ (Ohmic regime) and vanishes for $\beta \to \pi$ (tunneling regime).

One can now find the range of misalignment angles where a crossover from Ohmic to tunneling regime takes place. In terms of Fig.~\ref{fig:electric_current_diagram_maintext} this happens when the energy width of zone (I) becomes much larger than that of zones (II) and (III)
$$
eV \gg \frac{\hbar k_F}{m} \hbar\Delta k
\ \Rightarrow \
\frac{eV}{\epsilon_F} \gg \frac{\Delta k}{k_F} \ .
$$
Formula (\ref{eq:self-consistency}) then gives a condition for tunneling regime
\begin{equation}\label{eq:tunneling_regime_condition}
\frac{8(1/3 - \phi_2)}{\phi_1} \gg 1
\end{equation}
From this condition we find in Appendix~\ref{appsec:self-consistent_eV} that crossover from $eV/\epsilon_F = 0.1 \Delta k/k_F$ to $eV/\epsilon_F = 10 \Delta k/k_F$ happens in the interval $0.35 \, \beta_c < \beta < 1.15 \, \beta_c$, with Ohmic regime for smaller angles and tunneling for larger ones.

\section{Conclusions and discussion}\label{sec:conclusions_and_discussion}

It was always clear on general grounds \cite{slonczewski-tunnel:1989} that conductance of an F/F boundary with full spin polarizations of F-layers will vary between a large value in the parallel configuration (infinity in our model with matched bands), and zero in the antiparallel configuration. Our calculations show that when ferromagnetic gap $\Delta$ is much larger than Fermi energy $\epsilon_F$, crossover between the two limits happens around a very small angle $\beta_c \sim (\epsilon_F/\Delta)^{1/4}$ between the magnetizations, and has a width of order $\beta_c$. We also show that exchange interaction energy $E_{ex}(\beta)$ produced by itinerant electrons abruptly grows near $\beta_c$ from almost zero to almost its maximum value, producing a sharp peak of exchange torque located near $\beta_c$ and having a width of order $\beta_c$. This peak is in sharp contrast with the broad angular dependence found for conventional interlayer exchange in magnetic multilayers.

Phenomena described above are physically caused by efficient blocking of electron propagation through the boundary with increasing angle $\beta$. For an individual electron with perpendicular component of wave vector $k_L$, propagation switches from Ohmic to tunneling at an angle $\beta(k_L) \sim \sqrt{k_L/k_{\Delta}}$. When total conductance and total exchange torque are calculated, single-electron contributions are averaged with appropriate weight functions. Although the weights are not the same in two cases, it is still true that transition happens when most of electrons with $k_L \leq k_F$ are blocked from propagating through the boundary. This is how a common critical angle $\beta_c \sim \sqrt{k_F/k_{\Delta}} = (\epsilon_F/\Delta)^{1/4}$ emerges.

Our results show that crossover between Ohmic and tunneling regimes produces a region of nonlinear magnetic dynamics, where, for example, the peak of exchange torque may lead to a giant increase of ferromagnetic resonance frequency, and a rapid conductance drop can enable easy electronic readout. The newly found region is located in close proximity to the equilibrium parallel configuration, and thus can be reached easily and quickly. Its existence may lead to novel devices based on ferromagnetic materials with large values of exchange gap.

To discuss experimental realizations of the F/F devices suggested in this work, one needs to search for candidate materials and assess whether our simplified model's assumptions will apply to actual systems.

In Refs.~\cite{li:2020} and \cite{hao:2021} half-metallic ferromagnetism with $\Delta \gg \epsilon_F$ was theoretically predicted in 2D magnets. The values of the gap and Fermi energy were found to be of the order $\Delta \approx 1$~eV and $\epsilon_F \approx 0.1$~eV. Since our calculations neglect the temperature smearing of Fermi distribution, for them to be applicable one needs to require $k_B T \ll \epsilon_F$, which in this case translates into an experimentally reasonable condition $T \lesssim 100$~K.

One major process, known to be crucial in conventional spin valves but ignored in our theory, is the spin-flip scattering. In the bulk of half-metallic F-layers such scattering is suppressed in the following sense: since only up-spin states exist near the Fermi energy, the result of impurity scattering is not a propagating down-spin wave but a finite-size domain of down-spin around an impurity. The size of such domain is of the order $1/\kappa \approx 1/k_{\Delta}$. As long as impurities are located further than $1/k_{\Delta}$ away from each other, down-spin domains do not overlap and have no influence on spin transport in the bulk. The situation is, however, different for spin-flipping impurities that reside within the distance $1/k_{\Delta}$ from the boundary or in the ultrathin normal spacer discussed in the introduction. Spin flip on such impurities allows an electron to continue as a propagating wave on the other side of the boundary. While there may be few impurities in the layer of thickness $1/k_{\Delta}$, their presence becomes crucial at larger values of $\beta$, when clean surface conductance $G(\beta)$ drops to very small values. In this regime the total conductance $G_{tot} = G(\beta) + G_{imp}$ is dominated by the impurity spin-flip contribution and does not tend to infinity as predicted by our theory. Appendix~\ref{appsec:spin-flip_at_boundary} estimates $G_{imp}$ and shows that for low impurity concentration ``shunting by impurity scattering'' becomes appreciable only for $\beta \gg \beta_c$. Behavior near $\beta_c$ can therefore be well described by a model that neglects spin-flips. Condition of low impurity concentration states that the average distance between the impurities in the boundary layer has to be much larger than the Fermi wavelength $1/k_F$.

One can imagine other factors violating the assumptions of our theory. For example, will the spatial uniformity of magnetization in F-layers be violated due to the presence of thermal magnons? Analysis of this issue, as well as of other real-world effects, is left for the future.

\acknowledgements
We wish to thank Mark Stiles and Oleksiy Kolezhuk for illuminating discussions. We are grateful to the anonymous Referee whose insightful comments contributed to the enhancement of our paper.

\appendix 

\section{Single-electron currents through F/F boundary} \label{appsec:single-electron_currents}

\subsection{Transmission amplitudes}\label{app:transmission_amplitudes}

For a 3D wave vector $\bf k$ we define its parallel component ${\bf k}_{||}$ as a 2D vector parallel to the boundary plane. Perpendicular to the boundary component is denoted as $k_L$ for $x < 0$ and $k_R$ for $x > 0$.

Consider an $L \to R$ process with electron incoming from the left. The boundary between ferromagnets is assumed to be located at $x = 0$. The spinor of electron on the left of the boundary is
\begin{equation}\label{app:incoming_spinor_L-R}
\left( \begin{array}{c}
\psi_{\uparrow} \\ \psi_{\downarrow}
\end{array} \right)_L =
\left( \begin{array}{c}
e^{i k_L x} + r e^{-i k_L x} \\ d_L e^{\kappa_L x}
\end{array} \right) e^{i {\bf k}_{||} \cdot {\bf r}_{||}}, \quad (x < 0).
\end{equation}
On the right of the boundary the spinor is
\begin{equation}\label{app:transmitted_spinor_L-R}
\left( \begin{array}{c}
\psi_{\uparrow} \\ \psi_{\downarrow}
\end{array} \right)_R =
\left( \begin{array}{c}
t e^{i k_R x}  \\ d_R e^{-\kappa_R x}
\end{array} \right) e^{i {\bf k}_{||} \cdot {\bf r}_{||}},  \quad (x > 0).
\end{equation}
Here $r$ and $t$ are the reflection and transmission amplitudes, and $d_{L,R}$ are the amplitudes of decaying spin-down waves for an $L \to R$ process. Standard matching argument dictates equal ${\bf k}_{||}$'s on both sides. When electric current flows through an F/F structure at $\beta \neq 0$, a voltage drop is generally formed at the boundary (Fig.~\ref{fig:energy_diagram})---thus $k_{L}$ and $k_{R}$ may be different. Since we assume ferromagnets to be made from the same material with equal gaps $\Delta$ and aligned bottoms of energy bands at $V = 0$, the following relations are dictated by energy conservation in transmission process
\begin{eqnarray}
\label{app:kappa-k-delta}
  k_{L}^2 + \kappa_L^2 &=& k_{R}^2 + \kappa_R^2 = k_{\Delta}^2 \ ,
\\
\label{app:kL-kR}
  k_{L}^2 + k_V^2 &=& k_{R}^2.
\end{eqnarray}
where $k_{\Delta}^2 = 2 m \Delta/\hbar^2$, $k_V^2 = 2 m e V/\hbar^2$.

Spinors (\ref{app:incoming_spinor_L-R}) and (\ref{app:transmitted_spinor_L-R}) are written down with respect to two different quantization axis. On the left, $Z_L$ is directed along ${\bf M}_L$, and on the right $Z_R$ is directed along ${\bf M}_R$. Without loss of generality we will assume that ${\bf M}_R$ lies in the $(X_L,Z_L)$-plane (Fig.~\ref{fig:FF_device}). Matching of wave functions requires the use of appropriate spin rotation matrix
\begin{equation}\nonumber
\begin{gathered}
    R_y(\beta)=\begin{pmatrix}
            \cos{\beta/2} & -\sin{\beta/2}\\ \sin{\beta/2} & \cos{\beta/2}
        \end{pmatrix}.
\end{gathered}
\end{equation}
We then have a system of conditions to satisfy at $x = 0$
\begin{equation} \nonumber
    \begin{gathered}
        \psi_L = R_y \psi_R  \ , \\
        \frac{\partial\psi_L }{\partial x}  = R_y \frac{\partial\psi_R}{\partial x} \ .
        \label{boundary}
    \end{gathered}
\end{equation}
Using notation $s=\sin{\beta/2},c=\cos{\beta/2}$ we get a system of equations
\begin{equation}\nonumber
    \begin{gathered}
        1+r =t c- d_R s \ , \\
        d_L = d_R c+t s  \ , \\
        i k_L(1-r) = d_R \kappa_R s+i k_R t c  \ , \\
        d_L \kappa_L = -d_R \kappa_R c+i k_R t s  \ , \\
        \label{5}
    \end{gathered}
\end{equation}
from which we obtain expressions for reflection and transmission amplitudes
\begin{equation}\label{app:all_reflection_transmission_amplitudes}
    \begin{gathered}
        r=-\frac{(\kappa_R+i k_L) (\kappa_L-i k_R)s^2 -i  (\kappa_L+\kappa_R) (k_R-k_L)c^2}{D}, \\
        t=-\frac{2 i k_L  (\kappa_R+\kappa_L)c}{D}  \ , \\
        d_L=-\frac{2 i k_L  (\kappa_R+i k_R)s c}{D} \ , \\
        d_R=\frac{2 i k_L  (\kappa_L-i k_R)s}{D}   \\
    \end{gathered}
\end{equation}
with
\begin{eqnarray}
\nonumber
D &=&(\kappa_R - i k_L)(\kappa_L - i k_R) s^2 -
i (\kappa_L + \kappa_R)(k_R + k_L)c^2 .
\\
\label{app:D-s-c}
&&
\end{eqnarray}

\subsection{Approximations in the regime $k_V \ll k_F \ll k_{\Delta}$}

Amplitudes and single-electron current expressions can be simplified in the limit $\epsilon_F \ll \Delta$. We will also assume that the potential drop $eV$ at the boundary remains small compared to $\epsilon_F$. In terms of wave vectors this means
\begin{equation}\label{app:kv<<kF<<kD}
k_V \ll k_F \ll k_{\Delta} \ .
\end{equation}
Since $k_{L,R} \leq k_F$, strong inequality $k_{L,R} \ll k_{\Delta}$ holds for all electrons. On the other hand, perpendicular components $k_{L,R}$ of individual electrons' momenta can be either larger or smaller than $k_V$.

Denoting $k_R = k_L + \delta k$ and $\kappa_R = \kappa_L + \delta\kappa$, we find
\begin{eqnarray}
\delta\kappa &=& \sqrt{k_{\Delta}^2 - k_R^2} - \sqrt{k_{\Delta}^2 - k_L^2}
\approx -\frac{k_V^2}{2 k_{\Delta}} \ ,
\\
\delta k &=& \sqrt{k_L^2 + k_V^2} - k_L  \ .
\end{eqnarray}
Approximation in the first equation is always legitimate. In the second equation, approximation $\delta k \approx  k_V^2/(2 k_{\Delta})$ can be used only for $k_L \gg k_V$. Regardless of the relative values of $k_L$ and $k_V$, strong inequalities $\delta\kappa \ll \kappa_L$ and $\delta\kappa \ll \delta k$ always hold due to the double strong inequality (\ref{app:kv<<kF<<kD}). It is therefore possible to set
\begin{equation}\label{app:equal_kappa_approximation}
\kappa_L = \kappa_R = \sqrt{k_{\Delta}^2 - k_L^2} \approx k_{\Delta} -
\frac{k_L^2}{2 k_{\Delta}}
\end{equation}
in all expressions.

Approximate treatment of small $\delta k$ is more subtle and is discussed below. Let us demonstrate how it works in the case of denominator $D$ given by expression (\ref{eq:D-s-c}). Using Eq.~(\ref{app:equal_kappa_approximation}) we write
\begin{eqnarray*}
D & \approx & (\kappa - i k_L)(\kappa - i k_R) s^2 - 2 i \kappa (k_L + k_R)c^2
\\
&=& (\kappa^2 - k_L k_R) s^2 - i \kappa (k_L + k_R)(1+c^2) \ ,
\end{eqnarray*}
and therefore
\begin{eqnarray*}
&& |D|^2  \approx  (\kappa^2 - k_L k_R)^2 s^4 + \kappa^2 (k_L + k_R)^2(1+c^2)^2
\\
&& = k_{\Delta}^4 s^4 + k_{\Delta}^2\big[
  (k_L + k_R)^2(1+c^2)^2 - 2 k_L(k_L + k_R)s^4
\big]
\\
&& - k_L^2(k_L + k_R)^2 [(1+c^2)^2 - s^4] \ .
\end{eqnarray*}
We next use the smallness of $k_{L,R}/k_{\Delta} \ll 1$ to keep the terms of the orders of $k_{\Delta}^4$ and $k_{\Delta}^2 k_{R,L}^2$ but discard those of order $k_{L,R}^4$. That provides the next level of approximation
\begin{eqnarray}
\nonumber
|D|^2 & \approx &   k_{\Delta}^4 s^4 + k_{\Delta}^2
\big[  (k_L + k_R)^2(1+c^2)^2 -
\\
\nonumber
&&  - 2 k_L(k_L + k_R)s^4 \big]
\\
\label{app:|D|2_via_deltak}
&=& k_{\Delta}^4 s^4 + k_{\Delta}^2\big[
  (2 k_L + \delta k)^2(1+c^2)^2 -
\\
\nonumber
&&  - 2 k_L(2 k_L + \delta k)s^4 \big]
\\
\nonumber
&=& k_{\Delta}^2 \left\{ k_{\Delta}^2 s^4 +
4 k_L^2 \big[(1+c^2)^2 - s^4\big] + \right.
\\
\nonumber
&& + \left. (4 k_L \delta k + \delta k^2)(1+c^2)^2 - 2 k_L (\delta k) s^4
\right\} \ .
\end{eqnarray}
At this point we note that $\delta k$ satisfies an identity
$$
\delta k^2 = k_V^2 - 2 k_L\delta k
$$
and for $k_L \gg k_V$ admits approximations
\begin{eqnarray}
\nonumber
2 k_L \delta k & \approx & k_V^2 \quad (k_L \gg k_V),
\\
\label{app:approximations_for_delta-k}
{\delta k}^2 & \ll & k_V^2 \quad (k_L \gg k_V).
\end{eqnarray}
Anticipating integration of single-electron currents over a range of wave vectors that greatly exceeds $k_V$ (due to our assumption $k_F \gg k_V$), we conjecture that using these approximation will introduce a small relative error because they will be violated only in a small part of the integration interval.

With approximations (\ref{app:approximations_for_delta-k}) expression (\ref{app:|D|2_via_deltak}) for $|D|^2$ acquires a form
\begin{eqnarray}
\nonumber
|D|^2 & \approx &  k_{\Delta}^2 \left\{ k_{\Delta}^2 s^4 +
4 k_L^2 \big[(1+c^2)^2 - s^4\big] + \right.
\\
\nonumber
&& + \left. k_V^2 \big[2(1+c^2)^2 -  s^4 \big]
\right\}
\\
\label{app:|D|2_approximation}
&=& k_{\Delta}^2 \left\{ k_{\Delta}^2 s^4 +
16 k_L^2 c^2 + k_V^2 F(\beta) \right\}
\end{eqnarray}
with
\begin{equation}\label{app:definition_Fbeta}
F = 2(1+c^2)^2 -  s^4 = 1 + 6 c^2 + c^4 \ .
\end{equation}
Note a subtlety associated with expression (\ref{app:|D|2_approximation}): for any non-zero angle $\beta$ the second and third terms in $|D|^2$ become small compared to the first one for sufficiently small ratios $k_L/k_{\Delta}$ and $k_V/k_{\Delta}$. However, if we want to be able to vary $\beta$ from zero to $\pi$, we need to keep these terms to avoid a vanishing denominator at $\beta = 0$ (mathematically speaking, the limit of the denominator is non-uniform over the range of $\beta$).

We next apply the same approximation scheme to expression (\ref{eq:reflection_through_k's}) for reflection amplitude $r$. Using (\ref{app:equal_kappa_approximation}) we write
\begin{eqnarray}
\nonumber
r & \approx & - \frac{(\kappa^2 + k_L k_R)s^2 - i \kappa (k_L - k_R)(1+c^2)}{(\kappa^2 - k_L k_R) s^2 - i \kappa (k_L + k_R)(1+c^2)}
\end{eqnarray}
and
\begin{eqnarray}
\nonumber
|r|^2 & \approx & \frac{(\kappa^2 + k_L k_R)^2 s^4 + \kappa^2 {\delta k}^2 (1+c^2)^2}{(\kappa^2 - k_L k_R)^2 s^4 + \kappa^2 (2 k_L + {\delta k}^2)(1+c^2)^2} \ .
\end{eqnarray}
Discarding terms of order $(k_{L,R})^4$, and employing approximations (\ref{app:approximations_for_delta-k}), we further get
\begin{eqnarray}
\nonumber
|r|^2 & \approx & \frac{(k_{\Delta}^2 - k_V^2) s^4}{k_{\Delta}^2 s^4 +
16 k_L^2 c^2 + k_V^2 F(\beta)} \ .
\end{eqnarray}
In the denominator the term $k_V^2 F(\beta)$ can become comparable to $k_{\Delta}^2 s^4$ at $\beta \to 0$. Because of that, it cannot be discarded. In the numerator, $k_V^2$ can be safely neglected compared to $k_{\Delta}^2$. Our final result is
\begin{equation}\label{app:|r|2_approximation}
|r|^2  \approx  \frac{k_{\Delta}^2 s^4}{k_{\Delta}^2 s^4 +
16 k_L^2 c^2 + k_V^2 F(\beta)} \ .
\end{equation}

\begin{figure}[t]
\center
\includegraphics[width = 0.4\textwidth]{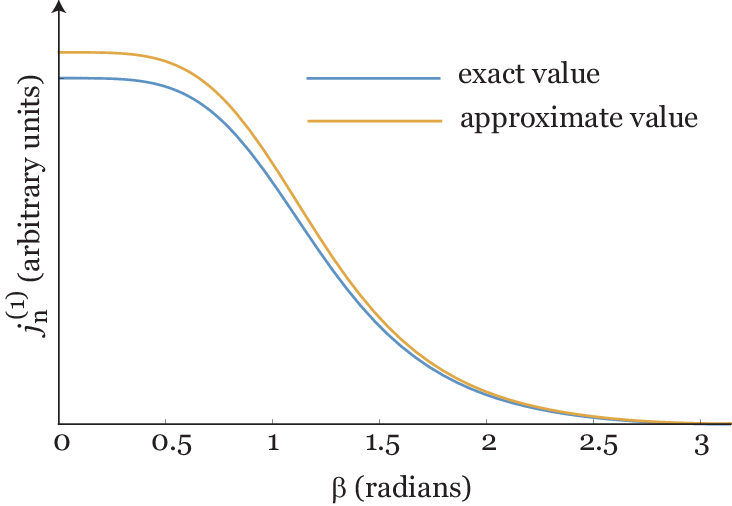}
\caption{Exact and approximate single-electron particle currents $j_{n}^{(1)}(\beta)$ for $k_L/k_{\Delta} = 0.1$, $k_V/k_L = 0.1$.}
 \label{fig:single_electron_jn_approximation_quality}
\end{figure}

One important lesson from the above is that expression for $r$ had to be expanded up to quadratic terms in $(k_{L,R}/k_{\Delta})^2$ in order to correctly expand $|r|^2$.

Single-electron particle current is given by formula (\ref{eq:je_through r}). In the approximation derived above we find
\begin{equation}\label{app:jnL}
j_{n}^{(1)} = \left( \frac{\hbar k_L}{m} \right)
\frac{16 k_L^2 c^2 + k_V^2 F(\beta)}{k_{\Delta}^2 s^4 + 16 k_L^2 c^2 + k_V^2 F(\beta)} \ .
\end{equation}
Figure~\ref{fig:single_electron_jn_approximation_quality} compares exact and approximate values of $j_{n}^{(1)}$ and demonstrates good quality of our approximation.

We could similarly work out an expression for the $R \to L$ current $j_{nR}^{(1)}$ but we will see below that it will not be necessary since all calculations for electric current can be done using the formula for $j_{nL}^{(1)}$ alone.

\begin{figure}[t]
\center
\includegraphics[width = 0.4\textwidth]{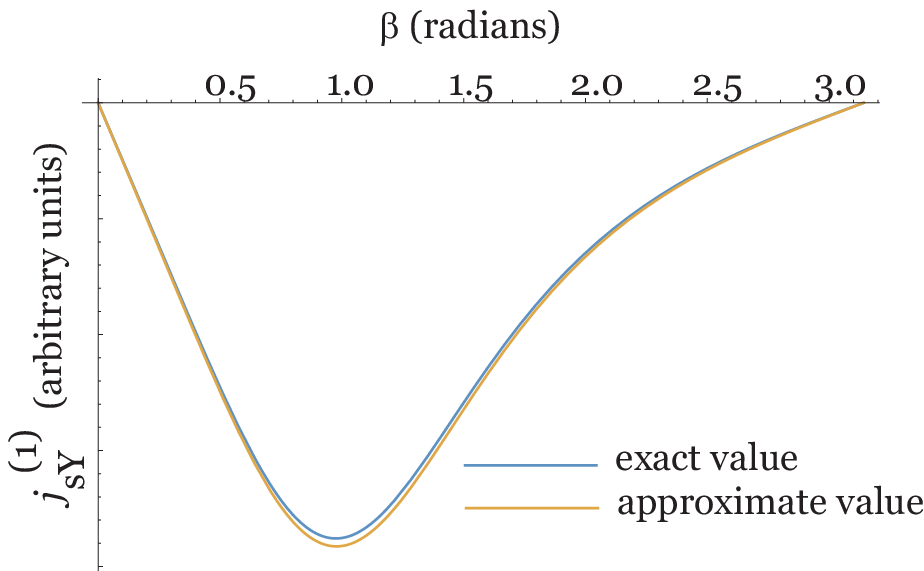}
\caption{Exact and approximate single-electron spin currents $j_{sY}^{(1)}(\beta)$ for $k_L/k_{\Delta} = 0.1$.}
 \label{fig:single_electron_js_approximation_quality}
\end{figure}

Single-electron spin current is given by formula (\ref{jY_through_r_and_dL}). Since our goal will be to calculate exchange torque in equilibrium, we are only going to need $j^{(1)}_{sY}$ at $V = 0$, where $\kappa_L = \kappa_R = \kappa$ and $k_L = k_R = k$. Moreover, as explained in the main text, $T_{eq}$ is expressed through the magnitude of spin current at $x = 0$ and equals
\begin{equation}\label{eq:js1_full}
j^{(1)}_{sY}(0) = - \frac{\hbar^2}{4 m} \big[ (\kappa + i k)(d_L + r d_L^*)  +  c. c. \big] \ .
\end{equation}
Current formula can be simplified due to the existence of another easily verifiable property
$$
(\kappa + i k) r d_L^* + c.c. = 0 \qquad (V = 0) \ ,
$$
that holds at $V =0$ for exact $r$ and $d$ given by expressions (\ref{app:all_reflection_transmission_amplitudes}). Consequently, formula (\ref{eq:js1_full}) reduces to
$$
j^{(1)}_{sY}(0) = -\frac{\hbar^2}{4 m} \big[ (\kappa + i k)d_L  +  c. c. \big]
\qquad (V = 0) \ .
$$
Amplitude $d_L$ is taken from Eq.~(\ref{app:all_reflection_transmission_amplitudes}) and at $V = 0$ equals
\begin{eqnarray*}
d_L & = & - \frac{2 i k (\kappa + i k) s \, c}{D_0} \ ,
\\
D_0 &=& (\kappa^2 - k^2) s^2 - 2 i \kappa k (1+c^2) \ .
\end{eqnarray*}
Therefore
$$
j^{(1)}_{sY}(0) =  -\frac{\hbar^2}{4 m} \left[
- \frac{2 i k (\kappa + i k)^2}{D_0}
+ \frac{2 i k (\kappa - i k)^2}{D^{*}_0}
\right] s \, c \ .
$$
Keeping terms up to $k^2$, we can reduce the above to
\begin{equation}\label{app:jsy_at_x=0}
j^{(1)}_{sY}(0) \approx - \left( \frac{\hbar^2}{4 m} \right) \frac{16 k^2 k_{\Delta} s \, c}{k_{\Delta}^2 s^4 + 16 k^2 c^2} \ .
\end{equation}
Figure~\ref{fig:single_electron_js_approximation_quality} compares exact and approximate values of $j_{n}^{(1)}$ and demonstrates good quality of our approximation.

\section{Relation between spin current and equilibrium exchange torque}
\label{appsec:jsY-TLY}
In our model spins of itinerant electrons experience exchange interaction with localized magnetic moments that form magnetization $\bf M$. This interaction is responsible for the exchange gap $\Delta$ in the energy spectrum. When an individual electron spin deviates from the direction of $\bf M$, a torque acting on it is generated. An opposite torque acts on the magnetization itself. After summation over itinerant electrons, one obtains the total torque ${\bf T}_s(x)$ acting on all of them at a given point in space. A torque ${\bf T}_m(x) = -{\bf T}_s(x)$ acts on magnetization at that same point.

Total angular momentum density $(\hbar/2)\bf S(x)$ of all itinerant electrons and total spin current satisfy the continuity equation \cite{bjz:1998, stiles_anatomyST:2002}
$$
\frac{\hbar}{2} \, \dot{\bf S}(x) + \frac{d{\bf j}_s}{dx} = {\bf T}_s(x) \ .
$$
In a stationary state, e.g. in equilibrium or with a d.c. electric current flowing, $\dot{\bf S}(x) = 0$. If we integrate the continuity equation over the extent of the left ferromagnet, $-\infty < x < 0$, we find
$$
{\bf j}_s(0) - {\bf j}_s(-\infty) = {\bf T}_s^{tot} \ ,
$$
where the right hand side is the total torque acting on all itinerant electrons present in the left F-layer.
Furthermore, ${\bf j}_s(-\infty)$ is directed along $Z_L$. This is a consequence of full spin polarization, or, in terms of spinors (\ref{app:incoming_spinor_L-R}), the result of the decay of down-spin components for each single-electron contribution at $x \to -\infty$. Therefore, for $Y$-components relation
$$
j_{sY}(0) = T_{sY}^{tot}
$$
holds in any stationary state, with or without a d.c. electric current. The final result for the torque acting on ${\bf M}_L$ is then
\begin{equation}\label{app:torque-spin-current-relation}
T_{LY} = - T_{sY}^{tot} = - j_{sY}(0) \ .
\end{equation}

The sign of $T_{LY}$ is of importance. When it is positive, the torque acts so as to push ${\bf M}_L$ towards ${\bf M_R}$ in Fig.~\ref{fig:FF_device}B. In equilibrium this corresponds to the whole system having energy minimum in the state with parallel magnetizations, i.e., to the ferromagnetic sign of exchange interaction. Negative value of $T_{LY}$ would mean energy maximum in the parallel configuration and energy minimum in the antiparallel state, i.e., an antiferromagnetic exchange.

Because the sign of $T_{LY}$ has direct measurable consequences, a natural question arises. What would have happened if majority electrons had their spins anti-parallel to $\bf M$? Such situation is realized in some materials. Could this possibly change the sign of $T_{LY}$ and switch the system from ferromagnetic to antiferromagnetic ground state? To answer this question, we repeat the wave function matching of Appendix~\ref{appsec:single-electron_currents} with spinors
\begin{equation}
\nonumber
\left( \begin{array}{c}
\psi_{\uparrow} \\ \psi_{\downarrow}
\end{array} \right)_L =
\left( \begin{array}{c}
d^A_L e^{\kappa_L x} \\ e^{i k_L x} + r^A e^{-i k_L x}
\end{array} \right) e^{i {\bf k}_{||} \cdot {\bf r}_{||}}, \quad (x < 0),
\end{equation}
and
\begin{equation}
\nonumber
\left( \begin{array}{c}
\psi_{\uparrow} \\ \psi_{\downarrow}
\end{array} \right)_R =
\left( \begin{array}{c}
d^A_R e^{-\kappa_R x} \\ t^A e^{i k_R x}
\end{array} \right) e^{i {\bf k}_{||} \cdot {\bf r}_{||}},  \quad (x > 0).
\end{equation}
where superscript ``$A$'' stands for ``antiparallel majority electrons''. Performing the calculations one finds exact relations $t^A = t$, $r^A = r$, $d^A_{L,R} = - d_{L,R}$. Using them in the spinors above, and substituting into the definition of spin current, we find, after a straightforward calculation
$$
j^A_{sY}(0) = j_{sY}(0) \ ,
$$
i.e., a material with majority electron spins antiparallel to $\bf M$ has the same sign of $T_{LY}$ as its counterpart with majority electron spins parallel to ${\bf M}$. Both materials should exhibit the same type of ground state.

\begin{figure}[t]
\center
\includegraphics[width = 0.16\textwidth]{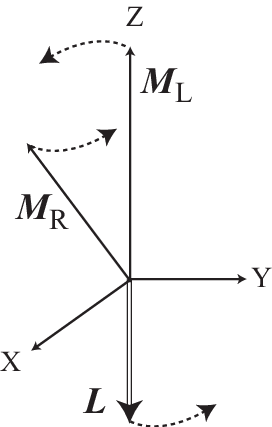}
\caption{Dynamic behavior of ${\bf M}_L$ is governed by the equation $\dot {\bf L} = {\bf T}_L$ for the angular momentum of the left layer, where ${\bf L} \uparrow \downarrow {\bf M}_L$. The sense of rotation is determined by $\dot L_Y = T_{LY} > 0$. Similar logic works for the dynamics of ${\bf M}_R$.}
 \label{fig:mutual_rotation_MR-ML}
\end{figure}

Our calculations are performed with an assumption of static magnetizations, held in place by external interactions, e.g., by strong anisotropies. Were magnetizations allowed to rotate, their dynamics would be governed by the equation $\dot {\bf L} = {\bf T}$, with ${\bf L}$ being the angular momentum associated with $\bf M$. Since $\bf L$ and $\bf M$ are antiparallel (gyromagnetic ratio is negative), the magnetizations would rotate around each other as depicted in Fig.~\ref{fig:mutual_rotation_MR-ML}, where the sense of rotation is shown for $T_{LY} > 0$.

\section{Approximations for equilibrium exchange torque}
\label{appsec:approximations_equilibrium_exchange}

\subsection{Position and height of the torque maximum}
Location of the maximum of $T_{LY}(\beta)$ can be determined from $d T_{LY}/d\beta = 0$, or equivalently from $d T_{LY}/ds = 0$. With angular dependence given by
$$
T_{LY} \sim \frac{s}{c} \, \phi_s(\nu_F)
$$
this leads to an equation
$$
\left( \frac{1}{c} + \frac{s^2}{c^3} \right) \phi_s + \frac{s}{c} \frac{d\phi_s}{d\nu_F} \frac{k_{\Delta}}{4 k_F} \left(\frac{2s}{c} + \frac{s^3}{c^3} \right) = 0 \ .
$$
Due to the presence of large factor $k_{\Delta}/k_F$ in the second term we know that solution $s_{max}$ will be close to zero. We can therefore approximate all quantities by their values at $s \to 0$, namely $c = 1$, $\phi_s = 1/3$, $d\phi_s/d\nu_F = - \pi/4$ and get an approximate equation
$$
\frac{1}{3} - \frac{\pi}{4} \frac{k_{\Delta}}{4 k_F} 2 s^2 = 0 \ ,
$$
which gives the maximum position
\begin{equation}\label{app:beta_max}
s_{max} = \sqrt{\frac{8 k_F}{3 \pi k_{\Delta}}} \ \Rightarrow \
\beta_{max} = 2 \sqrt{\frac{8 k_F}{3 \pi k_{\Delta}}}  \ .
\end{equation}
In terms of $\beta_c \approx 4 \sqrt{k_F/k_{\Delta}}$ that was defined in the main text,
$$
\beta_{max} = \sqrt{\frac{2}{3\pi}} \beta_c \approx 0.46 \beta_c \ .
$$

To calculate the value of  $T_{LY}$ at the peak's maximum, we first find
$$
\nu_{max} = \nu_F(\beta_{max}) \approx \frac{k_{\Delta}}{4 k_F} s_{max}^2
= \frac{2}{3\pi} \ .
$$
Then
\begin{eqnarray}
\nonumber
T_{LY}(\beta_{max}) &=& \left(\frac{\pi \hbar^2 k_{\Delta}k_F^3}{(2\pi)^3 m} \right)
s_{max} \ \phi_s(\nu_{max})
\\
\nonumber
&=& \left(\frac{\pi \hbar^2 k_{\Delta}k_F^3}{(2\pi)^3 m} \right)
\sqrt{\frac{8 k_F}{3 \pi k_{\Delta}}}
\ \phi_s\left( \frac{2}{3\pi} \right)
\\
\nonumber
&=& \left( \frac{\pi \hbar^2 k_F^4}{(2\pi)^3 m} \right) \sqrt{\frac{8 k_{\Delta}}{3 \pi k_{F}}} \ \phi_s\left( \frac{2}{3\pi} \right)
\\
\label{app:T_max}
& \approx & \left( \frac{\pi \hbar^2 k_F^4}{(2\pi)^3 m} \right)  0.19 \sqrt{\frac{k_{\Delta}}{k_{F}}}  \ .
\end{eqnarray}

\subsection{Total exchange energy}
We need to calculate
$$
E_{ex}^{tot} = \int_0^{\pi} T_{LY} d\beta = \left(\frac{\pi \hbar^2 k_{\Delta}k_F^3}{(2\pi)^3 m} \right) I
$$
with
\begin{equation}\label{app:exact_I}
I = \int_0^{\pi} \frac{s}{c} \, \phi_s(\nu_F(\beta)) d\beta \ .
\end{equation}
Changing integration variable to $x = \beta/2$, we re-write
$$
I = 2 \int_0^{\pi/2} \frac{\sin x}{\cos x} \phi_s \big( \nu_F(x) \big) dx \ ,
$$
with
\begin{equation}\label{app:nu_of x}
\nu_F(x) = \frac{k_{\Delta} \sin^2 x}{4 k_F \cos x} \ .
\end{equation}
Since
$$
d \nu_F = \frac{k_{\Delta}}{4 k_F} \sin x \left( 2 + \frac{\sin^2 x}{\cos^2 x} \right) dx \ ,
$$
the change of variable of integration to $\nu_F$ gives
\begin{eqnarray*}
I &=& 2 \int_0^{\infty} \frac{\sin x}{\cos x} \,
\frac{\phi_s(\nu_F) d\nu_F}{(k_{\Delta}/4 k_F) \sin x (2 + \sin^2 x/\cos^2 x)}
\\
&=& 2 \int_0^{\infty}
\frac{\phi_s(\nu_F) d\nu_F}{(k_{\Delta}/4 k_F) (2 \cos x + \sin^2 x/\cos x)}
\\
&=& 2 \int_0^{\infty} \frac{\phi_s(\nu_F) d\nu_F}{2 (k_{\Delta}/4 k_F) \cos x + \nu_F} .
\end{eqnarray*}
Inverting definition (\ref{app:nu_of x}) we find
$$
\cos x = \sqrt{\left( \frac{2 k_F \nu_F}{k_{\Delta}}\right)^2 + 1}
- \frac{2 k_F \nu_F}{k_{\Delta}} \ ,
$$
and therefore the denominator of the integrand simplifies to
$$
2 \frac{k_{\Delta}}{4 k_F} \cos x + \nu_F = \sqrt{\nu_F^2
+ \left( \frac{k_{\Delta}}{2 k_F}\right)^2} \ ,
$$
giving
$$
I = 2 \int_0^{\infty} \frac{\phi_s(\nu_F) d\nu_F}{\sqrt{\nu_F^2 + (k_{\Delta}/2 k_F)^2}} \ .
$$

So far all transformations were exact. Now we make an approximation based on $k_{\Delta}/k_F \gg 1$ and rapid decay of the function $\phi_s(\nu_F)$. The latter allows us to follow the usual three-step approximation process. First, we acknowledge that rapid decaly of $\phi_s$ allows us to replace the upper limit of integration by a number $A \gg 1$ without introducing a large error. Moreover, this number can be chosen so as to satisfy $1 \ll A \ll k_{\Delta}/2k_F$. Second, on the interval $(0,A)$ we can neglect $\nu_F^2$ in the denominator of the integrand. Finally, we again replace the upper limit by infinity, arguing that this operation will not introduce large error for the same reason as in the first step. We thus conclude that
\begin{equation}
I \approx 2 \int_0^{\infty} \frac{\phi_s(\nu_F) d\nu_F}{k_{\Delta}/2 k_F}
= \frac{4 k_F}{k_{\Delta}} \int_0^{\infty} \phi_s(\nu_F) d\nu_F \ .
\end{equation}
In our case the integral of $\phi_s$ can be calculated exactly \cite{computer_algebra}
$$
\int_0^{\infty} \phi_s(\nu_F) d\nu_F = \frac{\pi}{16} \ ,
$$
which gives
$$
I \approx \frac{\pi k_F}{4 k_{\Delta}}  \ .
$$
Note that we can estimate the neglected contribution from the interval $(A, \infty)$ by expanding $\phi_s$ in powers of $1/\nu_F$ at $\nu_F \gg 1$. The result is
\begin{eqnarray*}
&& \phi_s  \approx  \frac{1}{15 \nu_F^2} \ ,
\\
&& \frac{\phi_s(\nu_F)}{\sqrt{\nu_F^2 + (k_{\Delta}/2 k_F)^2}}  < \frac{2 k_F}{k_{\Delta}} \frac{1}{15 \nu_F^2}
\end{eqnarray*}
and
\begin{eqnarray*}
&& \Delta I = 2 \int_{A}^{\infty}
\frac{\phi_s(\nu_F) d\nu_F}{\sqrt{\nu_F^2 + (k_{\Delta}/2 k_F)^2}} <
\frac{4 k_F}{15 k_{\Delta}} \int_{A}^{\infty}\frac{d\nu_F}{\nu_F^2} =
\\
&& = \frac{4}{15} \frac{k_F}{k_{\Delta}} \frac{1}{A} \ll I \ ,
\end{eqnarray*}
where the last strong inequality is based on $A \gg 1$.

Being now confident in our approximation for integral $I$, we can write down the final result for exchange energy
\begin{equation}\label{app:Eex_approx_final}
E_{ex}^{tot} \approx \frac{\pi^2}{4(2\pi)^3}  \frac{\hbar^2 k_F^4}{m} \ .
\end{equation}

\section{Integrating single-electron contributions to electric current}
\label{appsec:electric_current_non_self-consistent}

Total particle current $j_n$ through the boundary depends on the distribution functions $n_L({\bf k})$ and $n_R({\bf k})$ of electrons reaching the boundary from left and right sides
$$
j_n = \int_{k_L > 0} j_{nL}^{(1)} \, n_L \frac{d^3 k}{(2\pi)^3} - \int_{k_R < 0} j_{nR}^{(1)} \, n_R \frac{d^3 k}{(2\pi)^3} \ .
$$
It is well known that in such calculations an identity holds
$$
j_{nL}^{(1)}({\bf k}_{||},k_L) \ \frac{d^2k_{||} d\epsilon}{v_{\perp L}} = j_{nR}^{(1)}({\bf k}_{||},k_R) \ \frac{d^2k_{||} d\epsilon}{v_{\perp R}} \ ,
$$
where $k_L$ and $k_R$ are related by $\epsilon({\bf k}_{||},k_L) = \epsilon({\bf k}_{||},k_R)$, and $v_{\perp} = \hbar k_{\perp}/m$ is the perpendicular to the boundary component of electron velocity. This means that the $L \to R$ current of electrons from a small element of momentum space is exactly compensated by the $R \to L$ current of electrons from an element of different size, corresponding to the same interval of energy. Because of this identity, full electric current can be expressed through $j_{nL}$ alone
\begin{eqnarray*}
j_n &=& \int_{k_L > 0} \frac{j_{nL}^{(1)}}{v_{\perp L}} \, \big[ n_L(k_L) - n_R(k_R(k_L)) \big] \frac{d^2 k_{||} d\epsilon}{(2\pi)^3} =
\\
&=& \int_{k_L > 0} j_{nL}^{(1)} \, \big[ n_L(k_L) - n_R(k_R) \big] \frac{d^3 k}{(2\pi)^3} \ .
\end{eqnarray*}
This property, in particular, ensures that electric current always vanishes in thermal equilibrium. In our situation, when $n_L$ and $n_R$ are equal to either zero or unity, it means that full current can be calculated as a sum of contributions of those $L \to R$ electrons that have no $R \to L$ counterparts with the same energy.

\begin{figure}[t]
\center
\includegraphics[width = 0.45\textwidth]{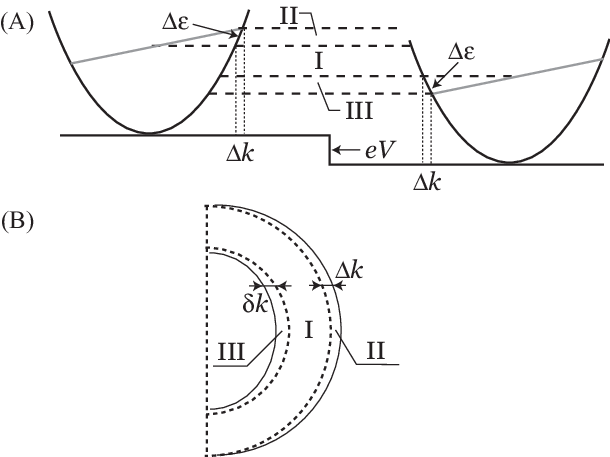}
\caption{Distributions of electrons on both sides of F/F boundary. Fermi surface shift by $\Delta k$ translates into the energy shift $\Delta\epsilon$. (A) Three intervals of energy contributing to electric current. (B) Regions in the $k$-space of the left F-layer, corresponding to the three energy intervals in (A). Dashed semi-circles with radii $k_F$ and $k_{(-)}$ define region I. Crescent-shaped regions II and III have widths $\Delta k$ and $\delta k$, see text.}
 \label{fig:electric_current_diagram}
\end{figure}

When both $n_L$ and $n_R$ are the ``shifted Fermi sphere'' distributions, three intervals of energy where $L \to R$ electrons have no $R \to L$ counterparts can be identified, as shown in Fig.~\ref{fig:electric_current_diagram}. First, there is a ``tunneling'' region (I) formed by electrons within the boundary voltage jump $eV$. Second, there are two ``Ohmic'' regions (II) and (III), related to the shift $\Delta k$. Their energy width is $\Delta \epsilon = \hbar^2 k_F \Delta k/m$. Here we start by expressing the contributions from regions (I), (II), and (III) through $eV$ and $\Delta k$, assuming that both parameters are known independently. In the actual stationary state with d.c. current there is a self-consistency connection between them, which we will derive and apply in Appendix~\ref{appsec:self-consistent_eV}.

\subsection{Tunneling contribution (region I)}
We use single-electron particle currents in approximation (\ref{app:jnL}) to find the sums. Tunneling contribution to the particle current is given by an integral
\begin{eqnarray}
\nonumber
j_{n(I)} &=& \frac{\hbar}{m} \int_{k_{(-)}}^{k_F} \frac{2\pi k^2}{(2 \pi)^3}
\int_0^1 j^{(1)}_{n}(\xi k) \, d\xi \, dk
\\
\nonumber
 &=& \frac{\hbar}{m} \int_{k_{(-)}}^{k_F} \frac{2\pi k^2 dk}{(2 \pi)^3} \times
\\
\nonumber
&& \times \int_0^1 d\xi
\frac{k \xi [16 k^2 c^2 \xi^2 + k_V^2 F(\beta)]}{k_{\Delta}^2 s^4 + 16 k^2 c^2 \xi^2 + k_V^2 F(\beta)} \ ,
\end{eqnarray}
where the lower limit is given by $k_{(-)}^2 = k_F^2 - k_V^2$. The angular part of the integral
$$
S_n = \int_0^1 d\xi
\frac{k \xi [16 k^2 c^2 \xi^2 + k_V^2 F(\beta)]}{k_{\Delta}^2 s^4 + 16 k^2 c^2 \xi^2 + k_V^2 F(\beta)}
$$
can be calculated by performing a variable change $t = \xi^2$. This results in
$$
S_n = \frac{k}{2} \left[
1 - \frac{k_{\Delta}^2 s^4}{16 k^2 c^2} \ln \left( 1 + \frac{16 k^2 c^2}{k_{\Delta}^2 s^4 + k_V^2 F(\beta)}  \right)
\right].
$$
Integration over $k$ is more complicated but if we assume the regime of small voltage drop $eV \ll \epsilon_F$, i.e., $k_V^2 \ll k_F^2$,  then $k_{(-)} \approx k_F$ and integration is performed over a narrow interval of width $k_F - k_{(-)} \approx k_V^2/2 k_F \ll k_F$. We therefore can approximate
$$
j_{n(I)} = \frac{\hbar}{m} \int_{k_{(-)}}^{k_F} \frac{2\pi k^2}{(2 \pi)^3} S_n dk
\approx
\frac{\hbar}{m} \frac{2\pi k_F^2}{(2 \pi)^3} S_n(k_F)(k_F - k_{(-)}) \ .
$$
Defining
\begin{eqnarray}
\label{app:dimensionless_gamma_and_chi}
\chi &=& \frac{k_V^2}{k_F^2} = \frac{eV}{\epsilon_F} \ ,
\\
\nonumber
\nu_F &=& \frac{k_{\Delta} s^2}{4 k_F c}
\end{eqnarray}
we re-write this result as
$$
j_{n(I)} = \frac{\pi}{2(2 \pi)^3}\frac{\hbar k_F^4}{m} \chi
\left[
1 - \nu_F^2 \ln \left( 1 + \frac{1}{\nu_F^2 + \chi F/16 c^2}  \right)
\right]
$$
or
\begin{eqnarray}
\nonumber
j_{n(I)} &=& \frac{\pi}{2(2 \pi)^3}\frac{\hbar k_F^4}{m} \chi \phi_1
\\
\label{app:jnI}
\phi_1 &=& 1 - \nu_F^2 \ln \left( 1 + \frac{1}{\nu_F^2 + \chi F/16 c^2}  \right) \ .
\end{eqnarray}
Since $\nu_F(\beta)$ becomes arbitrarily small at $\beta \to 0$, one cannot simplify say that $\nu_F^2 + \chi F/16 c^2 \approx \nu_F^2$ at $\chi \ll 1$. Nevertheless, it is possible to show that
\begin{equation}\label{app:phi_1_approximation}
\phi_1 \approx 1 - \nu_F^2 \ln \left( 1 + \frac{1}{\nu_F^2}  \right)
\end{equation}
is a very good approximation at small $\chi$.

\subsection{Ohmic contributions: regions (II) and (III)}
Region (II) is a crescent of constant thickness $\Delta k$ in $x$-direction. It has a circular left boundary with radius $k_F$. Due to the assumed smallness of $\Delta k/k_F$ we can express the contribution of this region as a surface integral over the half-sphere of radious $k_F$
$$
j_{n(II)} =\int_0^1 j^{(1)}_n(\xi k_F) \Delta k \, \xi \frac{2\pi k_F^2 d\xi}{(2\pi)^3}
$$
Using approximation (\ref{app:jnL}) for single-electron current, and notation (\ref{app:dimensionless_gamma_and_chi}), we get
\begin{equation}
\label{app:jnII_integral}
j_{n(II)} = \frac{2\pi}{(2\pi)^3} \frac{\hbar k_F^4}{m} \frac{\Delta k}{k_F}
\int_0^1 \frac{\xi^2(\xi^2 + \chi F/16 c^2)}{\nu_F^2 + \xi^2 + \chi F/16 c^2} d\xi \ .
\end{equation}
Evaluating the integral we get
\begin{eqnarray}
\nonumber
j_{n(II)} &=& \frac{2\pi}{(2\pi)^3} \frac{\hbar k_F^4}{m}
\frac{\Delta k}{k_F}  \phi_2 \ ,
\\
\label{app:phi2_definition}
\phi_2 &=& \frac{1}{3} - \nu_F^2 \left( 1 - \eta \arctan \frac{1}{\eta} \right) \ ,
\\
\nonumber
\eta &=& \sqrt{\nu_F^2 + \chi F/16 c^2} \ .
\end{eqnarray}
Again, it is possible to show that at $\chi \ll 1$
\begin{equation}\label{app:phi_2_approximation}
\phi_2 \approx \frac{1}{3} - \nu_F^2 \left( 1 - \nu_F \arctan \frac{1}{\nu_F} \right)
\end{equation}
is a very good approximation.

Region (III) is a crescent similar to that for region (II) with two modifications. First, it has a variable thickness $\delta k$ in $x$-direction, related to constant $\Delta k$ by the requirement of equal energy intervals
$$
k_L \delta k = k_R \Delta k \ .
$$
Since $k_R^2 = k_L^2 + k_V^2$, this gives
\begin{equation}\label{app:deltak}
\delta k = \frac{\sqrt{k_L^2 + k_V^2}}{k_L} \Delta k \ .
\end{equation}
Second, its circular right boundary has a radius $k_{(-)} = \sqrt{k_F^2 - k_V^2}$. Assuming small $\delta k$, we can again express the contribution of that region as a surface integral over a half-sphere of radius $k_{(-)}$
$$
j_{n(III)} = \int_0^1 j^{(1)}_n(\xi k_{(-)}) \, \delta k \, \xi \frac{2\pi k_{(-)}^2 d\xi}{(2\pi)^3} \ .
$$
Defining a dimensionless ratio
$$
q = \frac{k_{(-)}}{k_F} = \sqrt{1 - \chi} \ ,
$$
we re-write the current expression as
\begin{eqnarray*}
j_{n(III)} &=& \frac{2\pi}{(2\pi)^3} \frac{\hbar k_F^4}{m} \frac{\Delta k}{k_F} \times
\\
& \times & q^2 \int_0^1  \frac{\xi \sqrt{q^2 \xi^2 + \chi} \
(q^2 \xi^2 + \chi F/16 c^2)}{\nu_F^2 + q^2 \xi^2 + \chi F/16 c^2} d\xi \ .
\end{eqnarray*}
Performing a variable change $q \xi = \zeta$ we find
\begin{equation}\label{app:jnIII_integral}
j_{n(III)} = \frac{2\pi}{(2\pi)^3} \frac{\hbar k_F^4}{m} \frac{\Delta k}{k_F}
\int_0^{q}  \frac{\zeta \sqrt{\zeta^2 + \chi} \ (\zeta^2 + \chi F/16 c^2)}{\nu_F^2 + \zeta^2 + \chi F/16 c^2} d\zeta \ .
\end{equation}
This formula differs from expression (\ref{app:jnII_integral}) for $j_{n(II)}$  in two ways. The upper limit of integration is different, and the integrand has a factor $\zeta \sqrt{\zeta^2 + \chi}$ instead of $\xi^2$ in (\ref{app:jnII_integral}).

In the limit of interest $\chi \ll 1$ we can use approximations
\begin{eqnarray*}
&& \zeta \sqrt{\zeta^2 + \chi}  \approx  \zeta^2 + \frac{\chi}{2} \ ,
\\
&& q  \approx  1 - \frac{\chi}{2} \ ,
\end{eqnarray*}
and write
\begin{equation}\label{app:jnIII_integral_linear_in_chi}
j_{n(III)} = j_{n(II)} + \left( \frac{2\pi}{(2\pi)^3} \frac{\hbar k_F^4}{m} \frac{\Delta k}{k_F} \right) \frac{\chi}{2} \ \phi_3 \ ,
\end{equation}
with
\begin{eqnarray}
\nonumber
\phi_3 & = &
\int_0^{1}  \frac{\zeta^2 + \chi/2}{\nu_F^2 + \zeta^2 + \chi/2} d\zeta
- \frac{(1 + \chi/2) (1 + \chi/2)}{\nu_F^2 + 1 + \chi/2} =
\\
\nonumber
&=& 1 - \frac{\nu_F^2}{\sqrt{\nu_F^2 + \chi/2}} \arctan \left[ \frac{1}{\sqrt{\nu_F^2 + \chi/2}}  \right] -
\\
\label{app:phi3}
&-& \frac{(1 + \chi/2)^2}{\nu_F^2 + 1 + \chi/2}
\end{eqnarray}
When $\chi \ll 1$, the second term in formula (\ref{app:jnIII_integral_linear_in_chi}) will be small compared to the first, provided $\phi_3$ is not diverging or becoming too large. By plotting the graphs of $\phi_3$ we can numerically convince ourselves that it is bounded by
$$
|\phi_3| < 0.4 \ .
$$
Existence of this bound allows us to use the approximation $j_{n(III)} = j_{n(II)}$ for $\chi \ll 1$.

\subsection{Total current through the boundary}
Adding contributions from regions (I), (II), and (III), we obtain the total current through the boundary
\begin{eqnarray}
\nonumber
j_n &=&  j_{n(I)} + j_{n(II)} + j_{n(III)} =
\\
\nonumber
&=& \frac{1}{(2\pi)^3} \frac{\hbar k_F^4}{m}
\left[
\frac{\pi\phi_1}{2} \frac{eV}{\epsilon_F} + \left( 4\pi \phi_2 + 2\pi \frac{\chi}{2}\phi_3 \right)\frac{\Delta k}{k_F}
\right]
\\
\label{app:jn_total}
& \approx & \frac{1}{(2\pi)^3} \frac{\hbar k_F^4}{m}
\left[
\frac{\pi\phi_1}{2} \frac{eV}{\epsilon_F} + 4\pi \phi_2 \frac{\Delta k}{k_F}
\right]
\end{eqnarray}

\section{Self-consistent calculation of boundary voltage drop}
\label{appsec:self-consistent_eV}

\subsection{Current in the bulk}
Throughout this work we assume that the distribution function $n({\bf k})$ of electrons carrying electric current in the bulk is given by Botlzmann equation in isotropic scattering time approximation. A well known solution for this case is a filled Fermi sphere shifted in the direction of current propagation by a small wave vector $\Delta k = e E \tau/\hbar$, where $\tau$ is the relaxation time \cite{kittel_intro_solid_state, ashcroft-mermin_solid_state}. The total particle current corresponding to such distribution is
$$
j_n = \int \frac{\hbar k_x}{m} n({\bf k}) \frac{d^3k}{(2 \pi)^3} = \int \frac{\hbar k_x}{m} \big[ n({\bf k})-n_F({\bf k}) \big] \frac{d^3k}{(2 \pi)^3} \ ,
$$
where $n_F$ is the Fermi distribution (we consider low temperature $T \ll \epsilon_F$). Note a small difference from standard calculation: we have single occupancy of each $\bf k$-state, since only one direction of spin is allowed. The integral is calculated as
\begin{eqnarray}
\nonumber
j_n &=& \int_0^{\pi} \frac{\hbar k_F \cos\theta}{m}  (\Delta k \cos\theta)
\frac{2\pi  k_F^2 \sin\theta d\theta}{(2 \pi)^3} =
\\
\label{eq:jn_in_terms_of_dk/kF}
&=& \frac{4\pi}{3 (2 \pi)^3} \frac{\hbar k_F^3 \Delta k}{m} =  \frac{4\pi/3}{(2 \pi)^3} \frac{\hbar k_F^4}{m} \frac{\Delta k}{k_F} \ .
\end{eqnarray}

We can estimate the value of $\Delta k/k_F$ in real devices setting $e_F = 10$ eV for metallic ferromagnets and using the largest current density $j_e = 10^{12}$ A/m$^2$ observed in metallic spintronic devices. For these parameters
$$
k_F = \frac{\sqrt{2 m \epsilon_F}}{\hbar} \approx 1.6 \cdot 10^{10} \ 1/m
$$
and
$$
\frac{\Delta k}{k_F} = \frac{6 \pi^2 m}{\hbar k_F^4}\frac{j_{e}}{e} \approx 5 \cdot 10^{-5} \ .
$$

\subsection{Self-consistency condition}

Using relations (\ref{eq:jn_in_terms_of_dk/kF}) and (\ref{app:jn_total}), one can write down the condition of having the same electric current flowing in the bulk and through the boundary
$$
\frac{4\pi}{3 (2 \pi)^3} \frac{\hbar k_F^3 \Delta k}{m} =
\frac{1}{(2\pi)^3} \frac{\hbar k_F^4}{m}
\left[
\frac{\pi\phi_1}{2} \frac{eV}{\epsilon_F} + 4\pi \phi_2 \frac{\Delta k}{k_F}
\right]
$$
or
$$
4\pi \left( \frac{1}{3} - \phi_2 \right) \frac{\Delta k}{k_F} =
\frac{\pi \phi_1}{2} \frac{eV}{\epsilon_F} \ ,
$$
which finally gives
\begin{equation}\label{app:self-consistent_eV}
 \frac{eV}{\epsilon_F} = \frac{8 (1/3 - \phi_2)}{\phi_1} \frac{\Delta k}{k_F}
\end{equation}
with $\phi_{1,2}$ given by approximations (\ref{app:phi_1_approximation}) and (\ref{app:phi_2_approximation}).

\subsection{Boundary conductance}
Inverting Eq.~(\ref{app:self-consistent_eV})
$$
\frac{\Delta k}{k_F} = \frac{\phi_1}{8 (1/3 - \phi_2)} \frac{eV}{\epsilon_F} \ ,
$$
and substituting the result into Eq.~(\ref{eq:jn_in_terms_of_dk/kF}), we find for electric current
\begin{eqnarray*}
j_e &=& e j_n =  e \frac{4\pi/3}{(2 \pi)^3} \frac{\hbar k_F^4}{m}
\frac{\phi_1}{8 (1/3 - \phi_2)} \frac{eV}{\epsilon_F}
\\
&=& \frac{\pi}{(2 \pi)^3} \frac{e^2 k_F^2}{\hbar} \frac{\phi_1}{1 - 3\phi_2} V
= G(\beta) V
 \ ,
\end{eqnarray*}
where the conductance of the boundary is
\begin{eqnarray}
\label{app:boundary conductance}
G(\beta) &=& G_0 \frac{\phi_1}{1 - 3\phi_2} \ ,
\\
\label{app:G0}
G_0 &=& \frac{\pi}{(2 \pi)^3} \frac{e^2 k_F^2}{\hbar} \ .
\end{eqnarray}
As expected, conductance diverges at $\gamma \to 0$ (Ohmic regime) and vanishes at $\gamma \to \infty$ (tunneling regime).

\subsection{Crossover between Ohmic and tunneling-dominated regimes}
Tunneling regime is achieved when potential drop $eV$ on the boundary is large enough to ensure $eV \gg \Delta\epsilon = (\hbar k_F/m) \Delta k$, i.e., when the majority of contributing electrons reside in region I (Fig.~\ref{fig:electric_current_diagram}). In terms of wave vectors this requirement reads
\begin{equation}\label{eq:tunneling_regime_requirement}
k_V^2 \gg k_F \Delta k \quad \Rightarrow \quad \left(\frac{k_V}{k_F}\right)^2 \gg \frac{\Delta k}{k_F} \ ,
\end{equation}
or
\begin{equation}\label{eq:tunneling_regime_requirement_energy_form}
\chi = \frac{eV}{\epsilon_F} \gg  \frac{\Delta k}{k_F} \ .
\end{equation}
Using self-consistency relation (\ref{app:self-consistent_eV}), we find that tunneling regime is realized for
\begin{equation}\label{app:tunneling_regime_condition_on_beta}
\frac{8 (1/3 - \phi_2)}{\phi_1} \gg 1 \ .
\end{equation}

The left hand side of inequality (\ref{app:tunneling_regime_condition_on_beta}) is a function of $\nu_F$ alone. To estimate the width of the crossover region between Ohmic and tunneling regime, we numerically find the values of $\nu_{\pm}$ for which the l.h.s. is equal to 0.1 and 10. This gives an interval $(\nu_-,\nu_+) \approx (0.12, 1.35)$, or, equivalently, a range of misalignment angles
$$
0.35 \, \beta_c < \beta < 1.16 \, \beta_c
$$
expressed in terms of the characteristic angle $\beta_c = 4 \sqrt{k_F/k_{\Delta}}$. From that we conclude that $\beta_c$ is the angle around which a crossover from Ohmic to tunneling regime takes place. This transition is the root cause of rapid variations demonstrated by many properties of F/F system near $\beta_c$.

\begin{figure}[t]
\center
\includegraphics[width = 0.25\textwidth]{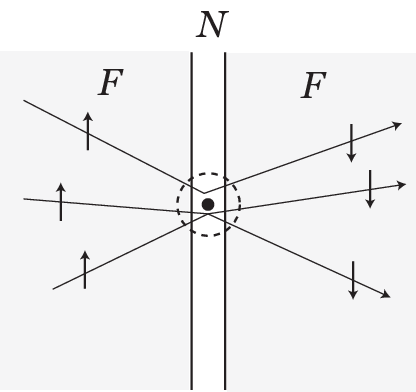}
\caption{Impurity at the boundary enables conductance by flipping electron spins.}
 \label{fig:impurity_in_N}
\end{figure}

\section{Effect of spin-flip scattering at the boundary} \label{appsec:spin-flip_at_boundary}
In this appendix we consider the effect of spin-flipping impurities located at the boundary separating the two ferromagnets in an F/F structure. Electrons colliding with such impurities can flip their spins (Fig.~\ref{fig:impurity_in_N}) and proceed to the other side even in the case of anti-parallel magnetic configuration where conventional propagation is completely blocked. As a result, transmission through impurities acts as a shunt inserted in parallel with the conductance considered in our study. Total conductance is a sum of clean boundary and impurities contributions
$$
G_{tot} = G(\beta) + G_{imp} \ ,
$$
with $G(\beta)$ is given by expression (\ref{app:boundary conductance}). We will assume that impurities have low concentration, so that their presence does not appreciably change $G(\beta)$. At the same time, when $G(\beta) \to 0$ the term $G_{imp}$ dominates in the total conductance. Here we want to estimate the angle $\beta_{shunt}$ at which the two terms become equal. For $\beta \gtrsim \beta_{shunt}$ our results for conductance, and possibly for the exchange torque, become questionable.

To produce a rough estimate, we will assume that after scattering off a spin-flipping impurity the electron can proceed into the other ferromanget without reflection. Clearly this will be the case of largest possible influence of impurities on the total boundary conductance. Then we can view each impurity as a quantum point contact between two ferromagnetic half-spaces (Fig.~\ref{fig:impurity_in_N}). Such point contact will contribute $e^2/(2\pi\hbar)$ to the total conductance of the boundary. Denoting the area concentration of impurities on the boundary as $n_{imp}^S$, we can write the impurity conductance per unit area as $G_{imp} = n_{imp}^S e^2/(2\pi\hbar)$. (If volume concentration $n_{imp}$ of spin-flip impurities is known and assumed to be constant throughout the structure, area concentration can be expressed through it: as argued in Sec.~\ref{sec:conclusions_and_discussion}, effective thickness of the boundary is $1/k_{\Delta}$, so $n_{imp}^S = n_{imp}/k_{\Delta}$). Equation $G(\beta) = G_{imp}$ then gives a condition
\begin{equation}\label{app:betashunt_equation}
\frac{\phi_1}{1 - 3\phi_2} = \frac{4\pi  n_{imp}^S}{k_F^2}
\end{equation}
where the left hand side is a function of $\beta$ according to the definitions (\ref{eq:phi1}), (\ref{eq:phi2}), and  (\ref{eq:nu_definition}). When $n_{imp}^S$ is small enough, so that the inequality
\begin{equation}\label{app:low_nimp_condition}
\frac{n_{imp}^S}{k_F^2} \ll 1
\end{equation}
holds, i.e., when the distance between impurities on the boundary is much larger than the electron Fermi-wavelength, the right hand side of Eq.~(\ref{app:betashunt_equation}) has to be small as well. The latter is only possible in the limit $\beta \gg \beta_c$, where the function on the left then can be approximated by
$$
\frac{\phi_1}{1 - 3\phi_2} \approx \frac{1}{2 \nu_F^2} \approx \frac{\beta_c^4}{2 \beta^4} \ .
$$
Equation (\ref{app:betashunt_equation}), when solved in this approximation, gives
$$
\beta_{shunt} = \beta_c \left( \frac{k_F^2}{8 \pi n_{imp}^S} \right)^{1/4} \gg \beta_c \ .
$$

In conclusion, we see that for the low impurity concentrations satisfying condition (\ref{app:low_nimp_condition}) the shunting effect will not modify our results until the magnetizations' misalignment angle increases well beyond the critical angle. All effects observed near the critical angle should be visible despite the wash-out due to impurity shunting. Note that inequality (\ref{app:low_nimp_condition}) also ensures that impurities can be viewed as independent quantum point contacts.

\end{document}